\documentclass[aps,pra,twocolumn,showpacs]{revtex4-1}

\usepackage{graphicx,amssymb,amsmath,color}

\begin{document}

\title{Precision lifetime measurement of the cesium $6P_{3/2}$ level using ultrafast pump-probe laser pulses}

\author{B. M. Patterson,$^{1}$ J. F. Sell,$^{1}$ T. Ehrenreich,$^{1}$ M. A. Gearba,$^{1,2}$ G. M. Brooke,$^{1}$ J. Scoville,$^{1}$ and R. J. Knize$^{1}$}

\affiliation{$^{1}$ Laser and Optics Research Center, Department of Physics, U. S. Air Force Academy, Colorado 80840, USA\\
$^{2}$ Department of Physics and Astronomy, University of Southern Mississippi, Hattiesburg, Mississippi 39406, USA}


\begin{abstract}

Using the inherent timing stability of pulses from a mode-locked laser, we have precisely measured the cesium $6P_{3/2}$ excited state lifetime.  An initial pump pulse excites cesium atoms in two counter-propagating atomic beams to the $6P_{3/2}$ level. A subsequent synchronized probe pulse ionizes atoms which remain in the excited state, and the photo-ions are collected and counted.  By selecting pump pulses which vary in time with respect to the probe pulses, we obtain a sampling of the excited state population in time, resulting in a lifetime value of 30.462(46)~ns. The measurement uncertainty (0.15\%) is larger than our previous report of 0.12\% [\textcolor{red}{Phys.~Rev.~A~\textbf{84}, 010501(R) (2011)}] due to the inclusion of additional data and systematic errors.  In this follow-up paper we present details of the primary systematic errors encountered in the measurement, which include atomic motion within the intensity profiles of the laser beams, quantum beating in the photo-ion signal, and radiation trapping.  Improvements to further reduce the experimental uncertainty are also discussed.

\end{abstract}


\maketitle

\section{Introduction}

Measurements of atomic lifetimes are useful in applications ranging from astrophysics \cite{Li2000} to laser design \cite{Beach2004}, while also providing fundamental insights into atomic structure.  Of particular importance, \textit{ab initio} calculations of atomic lifetimes rely on detailed knowledge of the relevant electronic wave functions \cite{Safronova1999, Gharibnejad2011}.  By comparing such calculations to precision lifetime measurements a direct test of the accuracy of the underlying theory is obtained.  Calculations are most accurate for alkali atoms because of their simple atomic structure, with uncertainties approaching 0.1\% in the case of cesium \cite{Porsev2010}.  We describe in detail our measurement of the Cs 6\textit{P}$_{3/2}$ excited state lifetime which achieves an experimental uncertainty of 0.15\%, compared to 0.23\% for the best previous direct measurement \cite{Rafac1999}.  This is accomplished by careful control of systematic errors along with a time base originating from pump and probe pulses from a mode-locked laser \cite{Patterson2003}.  Such lasers are already widely used in other areas of precision metrology as the basis for femtosecond frequency combs \cite{Jones2000} and the measurement of optical frequencies \cite{Udem1999}, and are used in ultrafast pump-probe spectroscopy \cite{Woutersen1997}.

A strong motivation for precision lifetime measurements in cesium is their importance in testing the reliability of models used to interpret atomic parity nonconservation (PNC) studies.  Atomic PNC measurements provide a unique test of the electroweak interaction in a regime that is complementary to high-energy accelerator techniques, and place constraints on new physics beyond the Standard Model \cite{Porsev2009,Davoudiasl2012}.  The interpretation of atomic PNC experiments requires accurate atomic structure calculations, such as the evaluation of radial matrix elements \cite{Dzuba2011}, which lifetime measurements directly provide.  Experimentally determined matrix elements, along with other parameters such as excitation energies and hyperfine splittings, can be compared to theoretically calculated values to test the precision of the underlying theory. Currently the most accurate atomic PNC measurement utilizes the parity-forbidden Cs $6S-7S$ transition \cite{Porsev2009, Wood1997, Bennett1999}.  Other atoms are also being pursued for PNC measurements such as Fr \cite{Sheng2010} and Ra$^{+}$ \cite{Wansbeek2008}, with a recent measurement demonstrating a large PNC effect in Ytterbium \cite{Tsigutkin2009}.  However, the required atomic structure calculations in Yb are much more complex than in Cs, preventing a test of the Standard Model.  Using our measurement of the 6\textit{P}$_{3/2}$ atomic state lifetime we directly determine the corresponding radial matrix element, providing a test of the calculated electronic wave functions used in interpreting cesium atomic PNC experiments.

The best agreement between theory and experiment in atomic lifetimes occurs in light atoms, with agreement at the 0.075\% level in He$^{+}$ \cite{Drake1992} where the atomic wave functions are exactly known, even approaching the 0.02\% level in Li where the theoretical uncertainty is stated to be $1 \times 10^{-6}$ \cite{Yan1995}.  Only recently has a comparison approaching the 0.1\% level in heavy alkali atoms become possible due to advances in relativistic many-body perturbation theory such as the all-order method \cite{Porsev2010, Safronova2007}.  Previous direct high-precision measurements of the Cs 6\textit{P}$_{3/2}$ lifetime employed position-correlated photon counting in a fast Cs beam with a reported value of 30.57(7) ns \cite{Rafac1999} and time-correlated single-photon counting in a thermal Cs beam resulting in a value of 30.41(10) ns \cite{Young1994}.  Indirect methods have also obtained precise lifetimes of the 6\textit{P}$_{3/2}$ state using the value of the van der Waals $C_{6}$ coefficient \cite{Derevianko2002} and from the Cs 6\textit{S}$_{1/2}$ static dipole polarizability \cite{Amini2003}.  Photoassociative molecular spectroscopy has also been used to extract atomic lifetimes with very small uncertainties, with a comprehensive review given by Bouloufa \textit{et al}. \cite{Bouloufa2009}.  However, as that reference discusses, while an earlier study obtained a very small uncertainty in the Cs 6\textit{P}$_{3/2}$ lifetime of 0.01\% \cite{Amiot2002}, a later re-analysis of the same data set resulted in an uncertainty of 1\% \cite{Bouloufa2007} necessitating the need for further measurements.  Other techniques such as beam-gas laser spectroscopy \cite{Volz1996} and high precision linewidth measurements \cite{Oates1996} have also achieved low uncertainty lifetime measurements in other atoms.  Using the technique described in this paper, we obtain a lifetime value of 30.462(46) ns, which allows a comparison between theory and experiment approaching the 0.1\% level in Cs. The uncertainty in the result (0.15\%) is somewhat larger than our earlier report \cite{Sell2011} of 0.12\% due to the inclusion of additional data sets and reassessment of the experimental errors. In this paper we address in detail the systematic errors of our technique.

\section{Experimental Technique}

\begin{figure}
\includegraphics[width=8.5cm]{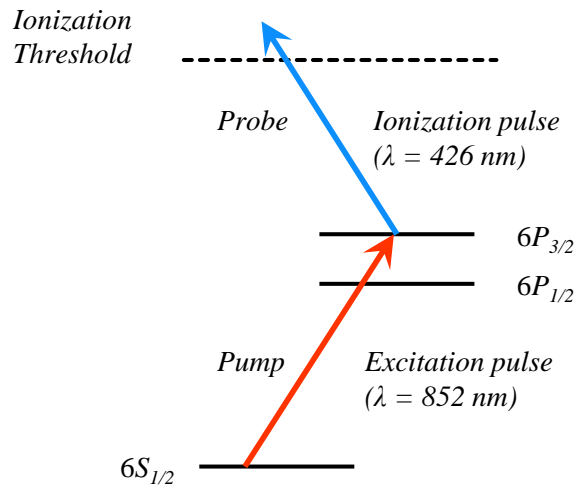}
\caption{\label{fig1-pump-probe} (Color online)  The pump-probe scheme.  A single mode-locked laser pulse ($\lambda _{pump}$ = 852 nm) excites cesium atoms to the 6\textit{P}$_{3/2}$ excited state.  A subsequent pulse is frequency doubled ($\lambda _{probe}$ = 426 nm) and ionizes atoms which remain in the excited state.}
\end{figure}

A simple pump-probe technique, illustrated in Fig.~\ref{fig1-pump-probe}, is used to precisely measure the Cs 6\textit{P}$_{3/2}$ radiative lifetime.  An initial laser pulse from a mode-locked laser excites Cs atoms to the 6\textit{P}$_{3/2}$ level, while a following synchronized probe pulse ionizes atoms which remain in the excited state.  The resulting photo-ions are collected using a Channeltron detector and counted.  This measurement process is repeated using different time delays between excitation and ionization pulses.  Since the photon energy in the ionization beam is sufficient to ionize atoms from the excited state but not from the ground state, the measurement effectively probes the excited state population as a function of time.  A simple exponential fit to the data points determines the radiative lifetime.  This technique has several attractive features.  First, the mode-locked pulse train provides a stable, intrinsic time base for the experiment.  We operate our ultrafast oscillator in a free running mode for this measurement, which gives the required timing accuracy; however, additional repetition rate and frequency stability can further be achieved by providing the necessary laser cavity feedback as used in stabilized frequency combs \cite{Cundiff2003}.  Second, each ionization pulse ionizes about 0.2\% of the excited atoms, which are then detected with near unity efficiency by the Channeltron. The total collection-detection efficiency (0.002) compares favorably to some fluorescence measurements, which collect only a fraction of emitted photons and may have collection-detection efficiencies of $10^{-4}$ or smaller (see, e.g., Ref.~\cite{Young1994}).  Lastly, the technique is quite general and can be applied to other atomic and molecular states, provided that the laser can supply the required excitation and ionization photon energies, and that the laser pulse repetition rate is relatively fast compared to the decay rate of the excited state of interest.

\begin{figure*}
\includegraphics[width=17cm]{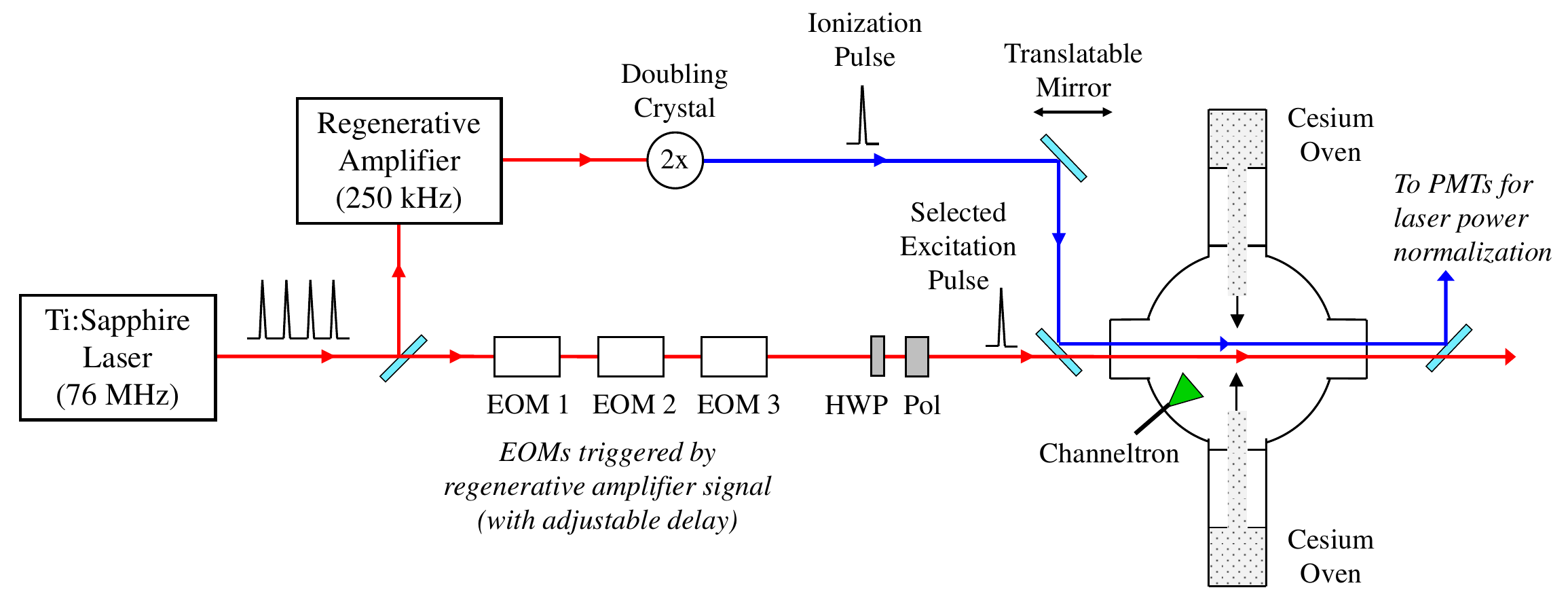}
\caption{\label{fig2-setup} (Color online)  The experimental setup.  A single mode-locked laser pulse ($\lambda _{pump}$ = 852 nm) excites cesium atoms to the 6\textit{P}$_{3/2}$ excited state.  A subsequent pulse is frequency doubled ($\lambda _{probe}$ = 426 nm) and ionizes excited state atoms. The photo-ions are collected with a Channeltron and counted. The polarization of the excitation beam is varied relative to the ionization beam using a computer-controlled polarizer (Pol) and half-wave plate (HWP). The collinear alignment of the two laser beams is optimized using a computer-controlled translatable mirror. The two laser beams exiting the chamber are separated and monitored with PMTs.}
\end{figure*}

Figure~\ref{fig2-setup} shows the experimental setup used to carry out the lifetime measurements.  Cesium atoms originate from counter-propagating and collimated thermal beams.  Atoms which enter the measurement region are excited to the 6\textit{P}$_{3/2}$ level by a single laser pulse selected from a mode-locked femtosecond laser using electro-optic modulators (EOMs).  A subsequent laser pulse from the same laser is amplified using a regenerative amplifier and then frequency doubled, ionizing atoms which are in the excited state.  Laser pulses originate from an ultrafast oscillator (Coherent Mira 900), which is a mode-locked Ti:Sapphire laser operating at a pulse repetition rate of 75.5~MHz and tuned to the $6S_{1/2} - 6P_{3/2}$ transition in cesium ($\lambda$ = 852 nm).  The laser pulses initially have a 150-fs pulse width, but are dispersively broadened to a few picoseconds by the EOMs. The pulse energy is approximately 0.8~nJ in the measurement region.  Due to the broad laser bandwidth ($\sim$8~nm FWHM), most of the photons are off-resonant with the atomic transition, with each laser pulse exciting only about 0.5\% of the cesium atoms that are illuminated. The fraction of atoms excited to the 6\textit{P}$_{1/2}$ state is negligible.  The ionization pulses are created by using a portion of the excitation laser output to seed a regenerative amplifier (Coherent RegA 9000) operating at 250~kHz.  The amplified pulses are subsequently frequency-doubled using a beta barium borate (BBO) crystal.  Each ionization pulse has a center wavelength of 426 nm and an energy of 1.3 $\mu$J, which ionizes approximately 0.2\% of excited 6\textit{P}$_{3/2}$ cesium atoms, assuming a photoionization cross-section of $1.2 \times 10^{-17}$ cm$^{2}$ \cite{Patterson1999}.  The excitation and ionization beams are combined using a dichroic mirror and aligned to be collinear.  A horizontal offset between the excitation and ionization laser beam axes can be introduced (or corrected for) using a computer-controlled translatable mirror (see Fig.~\ref{fig2-setup}).

\begin{figure}[b]
\includegraphics[width=8.5cm]{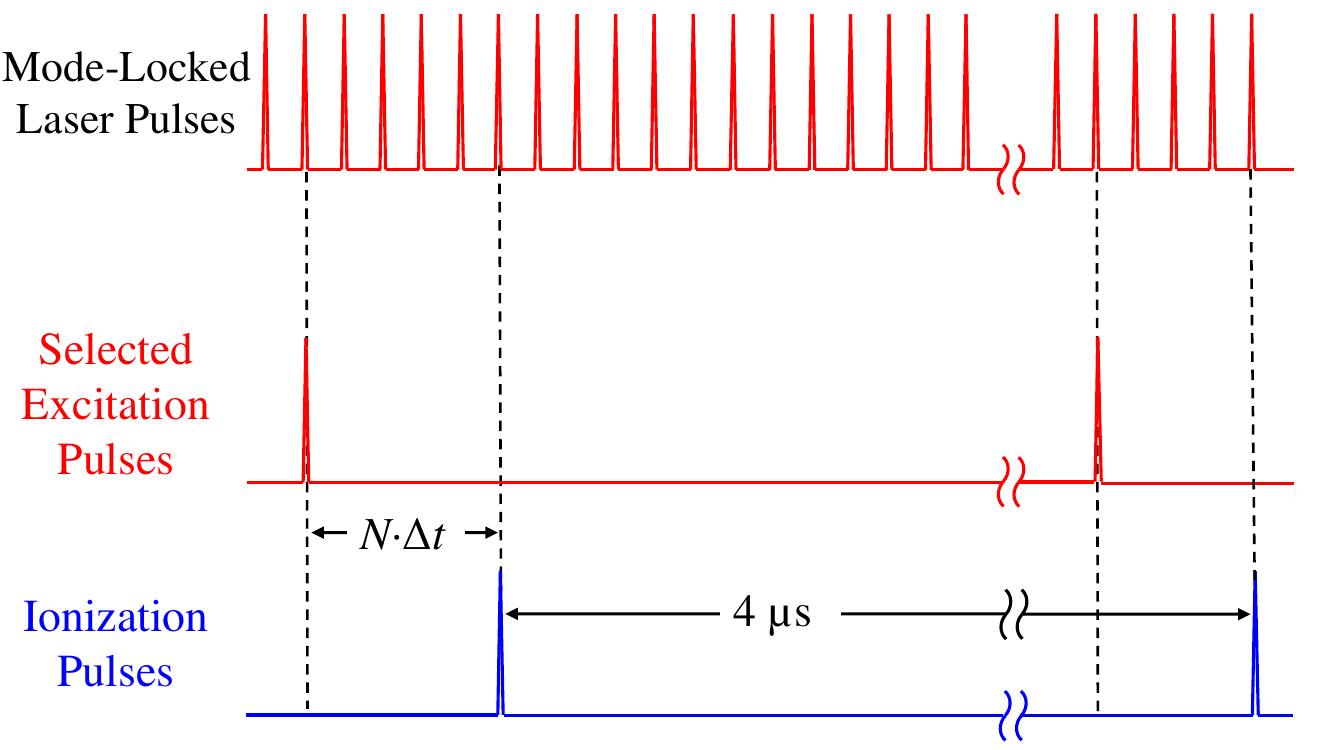}
\caption{\label{fig3-timing} (Color online)  Excitation and ionization laser pulse timing schematic.  The ionization pulses are seeded by the mode-locked laser and produced by a regenerative amplifier at a frequency of 250~kHz.  A synchronized signal from the regenerative amplifier triggers the EOMs to select excitation pulses, which are separated from the ionization pulses by an integral multiple of the mode-locked pulse interval $(N\cdot\Delta t)$.}
\end{figure}

The measurement technique achieves a high degree of timing accuracy by utilizing excitation pulses selected from a mode-locked pulse train, which are synchronized to the ionization pulses as illustrated schematically in Fig.~\ref{fig3-timing}.  Synchronization is achieved by having both the excitation and ionization pulses originate from the same ultrafast oscillator.  We choose excitation pulses which vary from the ionization pulses by a precisely known offset, $N \cdot \Delta t$, where $N$ is an integer and $\Delta t$ is the mode-locked pulse interval.  By choosing different excitation-ionization pulse pairs (i.e., by varying $N$), we obtain lifetime decay data separated in time by the laser pulse interval ($\Delta t \approx 13.24$ ns for this experiment).  The pulse interval was precisely determined to within 0.001\% during the measurements using a photodiode and a 225-MHz frequency counter (Agilent 53131A). There is an additional temporal offset between the excitation and ionization pulses due to different optical path lengths traversed, but this offset does not affect our measurements because it is constant for all pump-probe pulse pairs.  The excitation pulses are selected using three EOMs in series (two Conoptics models 360-80 and one Conoptics model 350-160) having a combined contrast ratio of better than $10^{5}$:1.  The EOMs are triggered using a synchronous signal from the regenerative amplifier at a frequency of 250~kHz.  The time delay between the ionization pulse and the triggering of the EOMs is adjusted using a pulse/delay generator (SRS DG535) to select a single excitation pulse.  Additional DG535 pulse/delay generators are used to fine tune the individual triggering of each EOM to the center of the selected excitation pulse.  The excitation-ionization pulse pair is repeated every 4 $\mu$s.

The ionization beam is linearly polarized in the horizontal direction, while the excitation beam polarization can be varied using a computer-controlled rotatable polarizer.  The half-wave plate shown in Fig.~\ref{fig2-setup} is used to maximize the intensity of the excitation beam for a selected polarization.  The angle $\theta$ between the two laser-beam polarizations is determined by mutually aligning each to a second polarizer positioned in the combined laser beams, thereby defining the zero angle ($\theta = 0^\circ$); this auxiliary polarizer is removed prior to the lifetime measurements.  The polarization alignment to the second polarizer is optimized by measuring the intensity of transmitted light while varying the polarizer angle in one-degree increments. These data were fit to a squared cosine function, according to Malus's law, to determine the zero angle.  With this scheme we can reliably control the relative polarization angle to within $\pm 0.25^\circ$.  For most measurements in the experiment, the relative polarization angle was set to the `magic angle,' $\theta = 54.7^\circ$, for which hyperfine quantum beats are suppressed.  Additional data were acquired for a range of polarization angles in order to study the effect of these quantum oscillations on the measured lifetime.

\begin{figure}
\includegraphics[width=8.5cm]{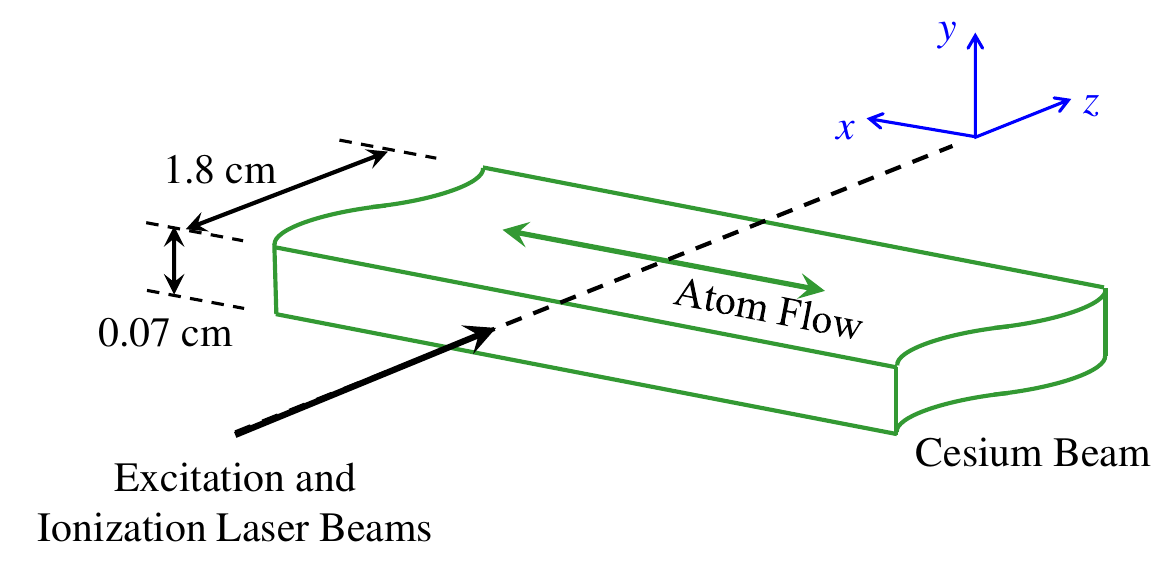}
\caption{\label{fig4-crossedbeams} (Color online)  The atomic and laser beam geometries.  The excitation and ionization laser beams intersect the cesium atomic beams at 90$^{\circ}$ and traverse it along its horizontal (1.8~cm) dimension.}
\end{figure}

The measurements are made in an ion-pumped vacuum system with a base pressure of $5 \times 10^{-9}$~Torr.  Separate ovens, temperature-controlled to within $\pm 0.1~^\circ$C, produce counter-propagating cesium atomic beams.  As discussed in detail later, the use of counter-propagating atoms mitigates systematic effects related to atomic motion within the spatially nonuniform laser beams.  Each cesium beam is collimated using two slits (1.7~cm $\times$ 0.0675~cm) separated by 55~cm, producing a beam with an approximate cross-section of 1.8~cm $\times$ 0.07~cm in the measurement region.  One of the beam dimensions is intentionally kept small to reduce the effects of radiation trapping.  As shown in Fig.~\ref{fig4-crossedbeams}, the excitation and ionization laser beams intersect the atomic beams perpendicularly along the broad (1.8-cm) dimension.  The excitation laser beam has a Gaussian spatial profile with a $1/e^{2}$ diameter of 0.10~cm.  The ionization beam has been expanded using a cylindrical lens and has an elliptical profile with $1/e^{2}$ diameters of 0.22~cm and 0.17~cm in the horizontal and vertical dimensions, respectively.  Upon exiting the chamber, the two laser beams are separated, attenuated, and used for power normalization of the photo-ion signal.  Both the excitation and ionization beams are continuously monitored during the measurements using photomultiplier tubes operating in photon-counting mode.  The photo-ions produced in the experiment are collected using a pulse-counting Channeltron detector (Burle 4800 series) and counted using a 225-MHz frequency counter (Agilent 53131A).

\begin{figure}[t]
\includegraphics[width=8.5cm]{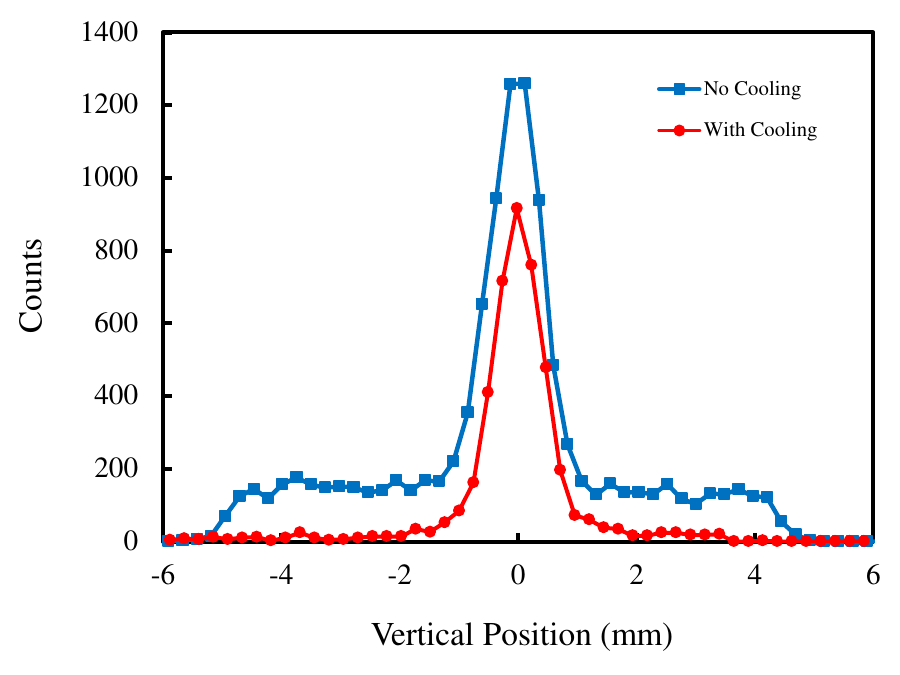}
\caption{\label{fig5-beamprofiles} (Color online)  The vertical profile of the counter-propagating cesium atomic beams. The data were acquired by sweeping the laser beams vertically through the atomic beams and counting the resulting photo-ions for a constant pump-probe delay time.  The atomic beamlines and collimation slits were maintained at room temperature (squares) and approximately $-40~^{\circ}$C (circles) during the measurements.}
\end{figure}

To achieve low-density atomic beams and reduce the effects of radiation trapping, the cesium ovens are maintained near room temperature (25.5~$^\circ$C) for most of the measurements considered here.  Some measurements are carried out at oven temperatures as high as 50~$^\circ$C to explore density dependent effects.  The cesium beamlines and collimation slits are cooled to $-40~^{\circ}$C to maintain well-defined atomic beams and reduce the Cs background density.  Figure~\ref{fig5-beamprofiles} shows a spatial profile of the cesium beam, acquired by simultaneously sweeping the excitation and ionization laser beams vertically through the 0.07-cm dimension of the atomic beam and counting the photo-ions produced, while keeping the pump-probe time delay constant.  The background cesium, which accounts for about 10\% of the ion signal at room temperature, is largely eliminated by cooling the beamlines and slits.

\begin{figure}[b]
\includegraphics[width=8.5cm]{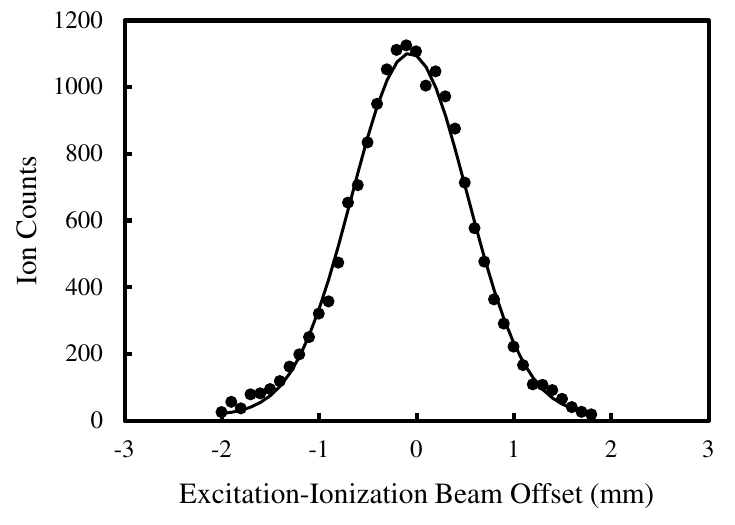}
\caption{\label{fig6-laseroffset}  The photo-ion signal produced while translating the ionization laser beam horizontally through the excitation laser beam for a constant pump-probe delay time.  The solid line is a Gaussian fit to the data.  The ionization beam is positioned at the peak of the curve to optimize the alignment of the two laser beams before each data run.}
\end{figure}
The collinear alignment of the two laser beams is optimized prior to each data run by translating the ionization beam horizontally through the excitation beam in 0.1-mm increments, while counting the ions produced for a constant pump-probe delay time.  Because the photo-ion signal is proportional to the spatial overlap of the two laser beams, this measurement yields the convolution of the laser beam profiles.  The photo-ion signal is shown in Fig.~\ref{fig6-laseroffset} as a function of the beam offset and is well described by a Gaussian function, which is the expected convolution for the two Gaussian laser beams.  The reduced chi-squared value for the central region of the data (from -1.0 to +1.0~mm offset), which is the interval used for peak determination, is 1.3.  There is increased scatter in the wings of the distribution, which has no significant effect on the measurement. The ionization beam is positioned at the peak of the fit curve, which can be achieved to within $\pm 20~\mu$m.  This uncertainty is small compared to the spatial drift during a measurement, so the beam overlap is re-optimized following each data run.  The magnitude of the observed positional shift is typically less than 100~$\mu$m, which we take as the uncertainty in the horizontal beam alignment.  The vertical beam alignment is not critical, as will be discussed, because it is perpendicular to the atomic flow direction.

\section{Results}

Data were acquired for delay times ranging from 0 to 530~ns (0 to 40 mode-locked pulse intervals).  Ion counts were typically accumulated for three seconds at each delay, for a total count time of two minutes required to produce a full decay curve. Photon counts from the two photomultiplier tubes used for laser power normalization (see Fig.~\ref{fig2-setup}) were also recorded for each pulse delay. This sequence was repeated 20 times in a typical data run for an accumulated count time of forty minutes. A few data sets were acquired using a slightly higher count time (5 seconds) while maintaining a total accumulation time of about 40 minutes. Accumulating the ion counts in small time bins helps to average out the effects of any drifts in the power and pulse interval of the mode-locked laser and the regenerative amplifier.  The collinear alignment of the excitation and ionization beams was optimized before and after each 40-minute data run and the amount of beam drift observed during the run was recorded.  The mode-locked pulse interval $\Delta t$ was also precisely measured before and after each run, as this value is critical in determining the time scale for the decay.

\begin{figure}
\includegraphics[width=8.5cm]{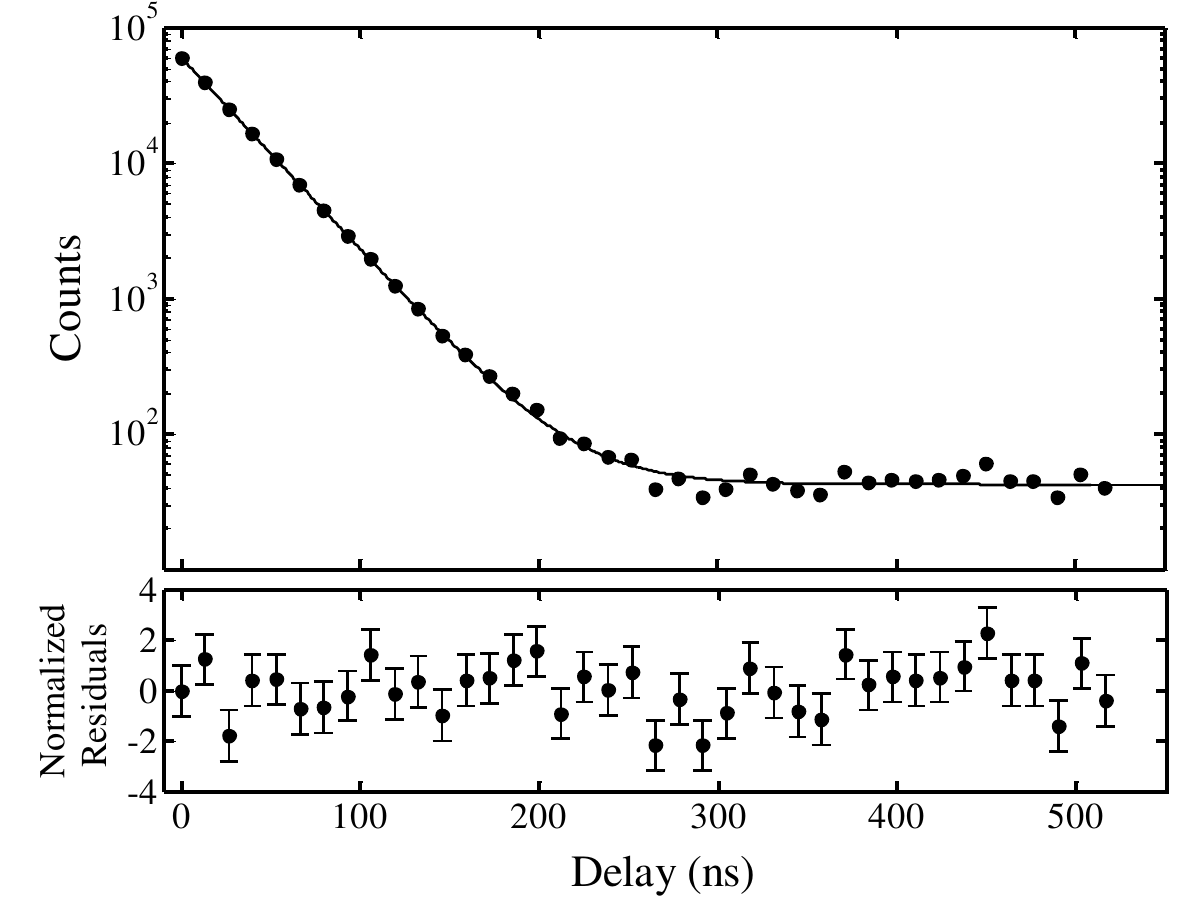}
\caption{\label{fig7-sampledata}  Total accumulated ion counts for twenty individual two-minute lifetime measurements (one 40-minute data run). The data were acquired at a cesium oven temperature of 25.5~$^\circ$C. The solid line is a fit to the data using Eq.~1 and the fit residuals are normalized in units of the ion count uncertainty.}
\end{figure}

\begin{figure}[b]
\includegraphics[width=8.5cm]{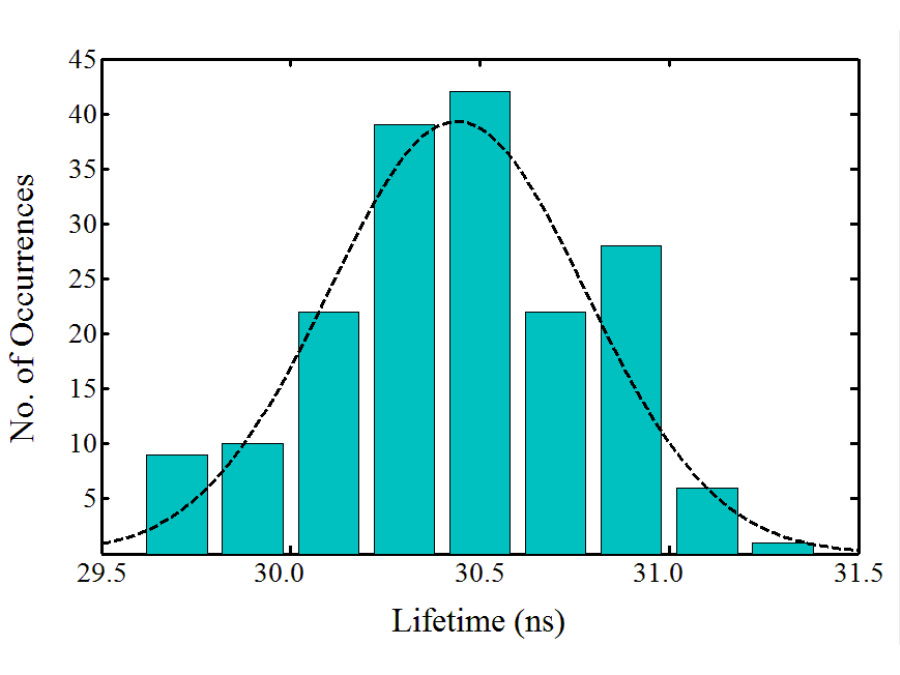}
\caption{\label{fig8-histogram}  (Color online) Histogram of 180 individual lifetime measurements with a mean lifetime of 30.445~ns and a statistical uncertainty of 0.022~ns.  The dashed line is a Gaussian curve illustrating the mean and standard deviation of the measurements, and fits the data with a reduced chi-squared value of 1.2.}
\end{figure}

The ion counts were corrected for laser power drifts by dividing each recorded ion count by the product of the two PMT photon-count signals in the corresponding time bin, and multiplying by an overall constant to preserve the total number of counts for the data run.  The corrected data for each two-minute decay curve were fit to an exponential function with a constant background term,
\begin{equation}
    y(t)=Ae^{-t/\tau}+B
\end{equation}
 where $\tau$ is the lifetime, $A$ is the initial ion count, and $B$ is the background.  To better reveal the presence of any systematic variation in the fit residuals, we also fit the total accumulated counts for each 40-minute data run. An example of the latter fit is shown in Fig.~\ref{fig7-sampledata}, where no sign of a systematic variation is observed in the fit residuals. The reduced chi-squared value for this fit is 1.1, indicating the data are statistically consistent with an exponential decay. A background signal of $\sim$0.1\% is seen in the decay curve and is typical of these measurements. It is attributed to residual excitation caused by unselected laser pulses which are not fully extinguished by the EOMs. Because we are selecting fewer than one excitation pulse out of every 100 mode-locked laser pulses, even the high EOM contrast ratio ($10^{5}$:1) would allow background excitation at a relative level of about $100\cdot10^{-5}$ or 0.1\% compared to the selected pulse. Finally, the histogram shown in Fig.~\ref{fig8-histogram} compiles the results of 180 individual two-minute lifetime measurements, which were acquired at a cesium oven temperature of 25.5~$^\circ$C and a relative polarization angle of $\theta = 54.7^\circ$ between the excitation and ionization laser beams.  These measurements have a weighted mean of 30.445~ns with a statistical uncertainty of 0.022~ns (0.07\%).  The average lifetime was subsequently corrected for the effects of radiation trapping and for a systematic effect arising from the spatially nonuniform laser beam intensities, as discussed below. Our final corrected lifetime value is 30.462~ns with an overall uncertainty of 0.046~ns (0.15\%).

\section{Discussion of Systematic Effects}

The dominant systematic errors in this experiment, summarized in Table~I, include the consequences of the spatial non-uniformity and relative alignment of the excitation and ionization beams; hyperfine and Zeeman quantum beats; and, at elevated temperatures, radiation trapping. The remaining systematic errors analyzed for this experiment are small ($\leq$ 0.005\%) and include the effects of the timing stability of the mode-locked pulse train; imperfect EOM pulse selection; and pulse pileup in the Channeltron signal.

\begin{table}[h]
\caption{\label{tab:table1}   Contributions to the uncertainty in the cesium $6P_{3/2}$ state lifetime.}
\begin{ruledtabular}
\begin{tabular}{p{4.5cm} c c}
Source                                                &Correction ($\%$) & Error ($\%$) \\
\hline
Laser beam offset:                                    &            &               \\
\hspace{1em}(a) atom flux imbalance                   &            & $\pm$0.05    \\
\hspace{1em}(b) second-order correction               & $+0.18$    & $\pm$0.02      \medskip\\
Hyperfine quantum beats:                              &            &               \\
\hspace{1em}(a) polarization angle $\theta$           &            & $\pm$0.06    \\
\hspace{1em}(b) circular polarization                 &            & $\pm$0.04      \medskip\\
Zeeman quantum beats                                  &            & $\leq$ 0.01   \\
Mode-locked pulse stability                           &            & $\leq$ 0.001  \\
EOM pulse selection                                   &            & $\leq$ 0.005  \\
Pulse pileup                                          &            & $\leq$ 0.002   \medskip\\
Statistics and extrapolation to zero Cs density       &            & $\pm$0.12    \\
\hline
Total                                                 &            & $\pm$0.15    \\
\end{tabular}
\end{ruledtabular}
\end{table}

\subsection{Radiation Trapping}
\begin{figure}
\includegraphics[width=8.5cm]{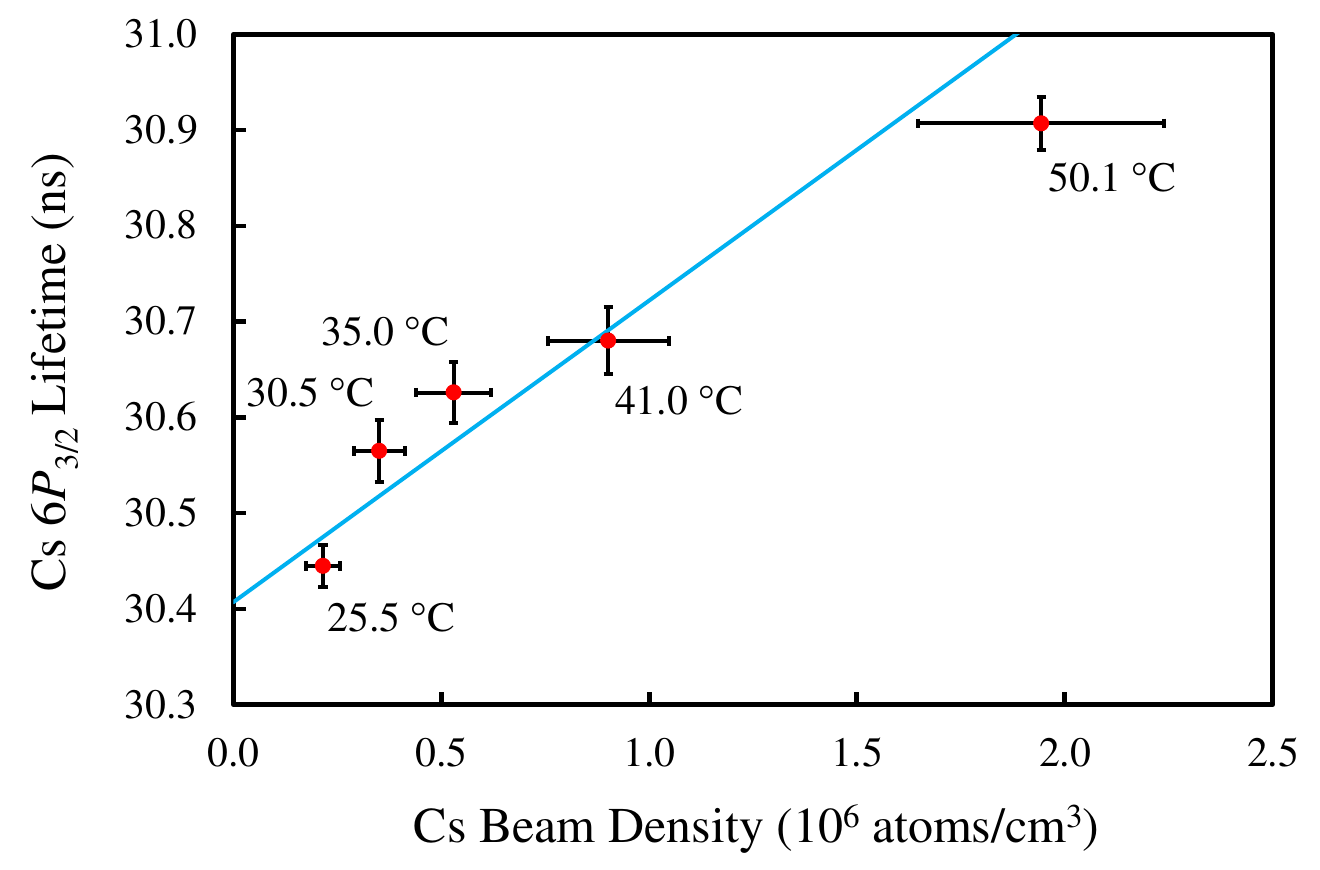}
\caption{\label{fig9-radiationtrapping}  (Color online) The measured $6P_{3/2}$ state lifetimes for a range of Cs oven temperatures. The solid line is a weighted linear fit to the data and yields an extrapolated lifetime of 30.407~ns for zero Cs density, with a statistical uncertainty of 0.038~ns.}
\end{figure}

In sufficiently dense atomic beams, the measured lifetime is increased due to radiation trapping and other density dependent effects. We characterized these effects by measuring the Cs $6P_{3/2}$ state lifetime for a range of oven temperatures, as shown in Fig.~\ref{fig9-radiationtrapping}. Each point on this graph represents the average of 40 to 120 individual lifetime measurements, with more data required for lower oven temperatures (where the count rates are reduced) to achieve the same counting statistics. For each oven temperature, the corresponding Cs beam density (in m$^{-3}$) was determined from gas kinetic considerations using \cite{Ramsey1956}
\begin{equation}
    n = 2\alpha \frac{A_{s}A_{c}}{L^{2}A} \frac{P}{T}
\end{equation}
where $P$ and $T$ are the Cs vapor pressure and absolute temperature inside the oven;  $A_{s}$ and $A_{c}$ are the areas of the oven source slit and the collimation slit, respectively, and $L$ is their separation; and $A$ is the cross-sectional area of the atomic beam in the probe region.  The constant $\alpha$ has a value of $4.892\times10^{15}$~K/Pa$\cdot$s if the other parameters are given in SI units, and the overall factor of two accounts for having two ovens. The density uncertainty is based on the uncertainty in the oven temperature, which we take as the spatial variation of the temperature ($\sim2~^\circ$C) over the oven's exterior. The data in Fig.~\ref{fig9-radiationtrapping} were fit to a straight line, $y = a + bx$, with the uncertainties in both density and lifetime accounted for by minimizing chi-squared in the form
\begin{eqnarray}
    \chi^{2} = \sum_{i=1}^{n} \frac{ \left( y_{i} - a - b x_{i} \right)^{2} }{ {\sigma_{yi}}^2 + b^{2} {\sigma_{xi}}^2 }.
\end{eqnarray}
The uncertainties in the fit parameters were determined using a Monte Carlo approach to simulate and fit random data sets, gaussian-distributed about the original data.   The fit gives an extrapolated lifetime value of 30.407~ns at zero Cs density, with a statistical uncertainty of 0.038~ns (0.12\%). This corresponds to only a slight (-0.12\%) correction to the data obtained at $25.5^\circ$C.

We attribute the lifetime shifts shown in Fig.~\ref{fig9-radiationtrapping} to radiation trapping, although the size of the effect is about an order of magnitude too large when compared to a simple estimate based on the atomic beam dimensions and temperatures used. In principle, the discrepancy could be due to background cesium, but we have largely eliminated this possibility using cryogenic cooling (Fig.~5). Increased pulse pileup in the Channeltron at higher cesium densities is another possible explanation, but this effect is small (0.002\%) for our experiment, as will be discussed later. It is most likely that the calculated cesium densities underestimate the actual beam densities, which were not directly measured in this experiment. Measurement of the absolute densities is not critical, however, as long as the relative densities are known. That is, the extrapolated lifetime value in Fig.~\ref{fig9-radiationtrapping} is accurate, assuming the calculated densities disagree with the actual beam densities by at most a constant factor. Finally, it is possible that our estimate of the radiation trapping effect on the lifetime is overly simplistic; we are therefore conducting more rigorous Monte Carlo calculations to better characterize this effect.

\subsection{Offset of Excitation and Ionization Beams}

\begin{figure}
\includegraphics[width=8.5cm]{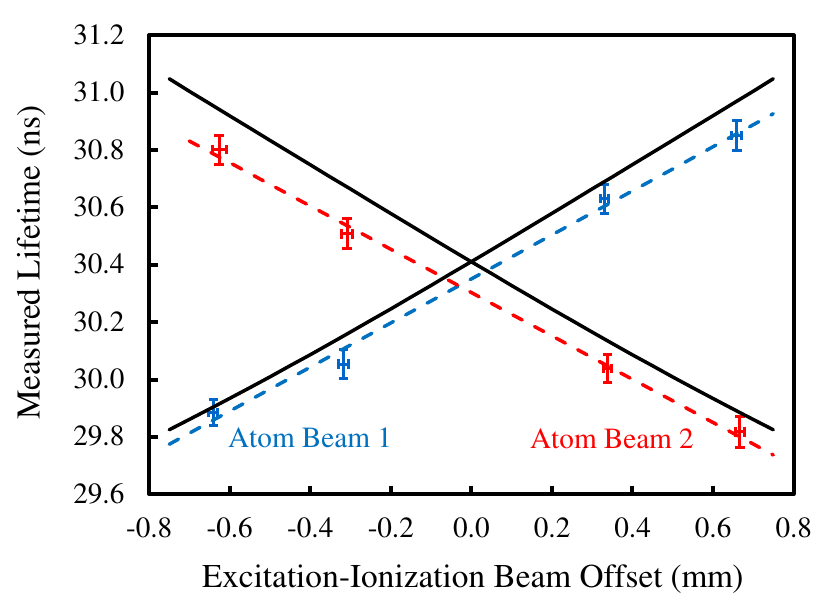}
\caption{\label{fig10-beamoffset}  (Color online) The Cs~6\textit{P}$_{3/2}$ lifetime as a function of the horizontal offset between the pump and probe laser beams, determined separately for each counter-propagating atomic beam. The isolated data points correspond to experimental measurements and the solid lines are model predictions.}
\end{figure}

A systematic effect arises from the motion of excited atoms through the nonuniform intensity profile of the probe (ionization) beam. An atom is exposed to a greater or lesser ionizing intensity depending upon its position when the probe beam is introduced. Consequently the detection efficiency is not uniform but varies systematically with the delay time, causing a measurable shift in the experimental lifetime. A similar uncertainty arises from the relative alignment of the pump and probe beams. To assess these effects experimentally, we deliberately introduced a small horizontal offset between the two laser beam axes as measured along the atom flow direction using the translatable mirror shown in Fig.~\ref{fig2-setup}. The measured lifetime is plotted as a function of this offset in Fig.~\ref{fig10-beamoffset}. The data were acquired separately for each of the two counter-propagating atomic beams and yielded a lifetime shift of approximately 0.76(4)~ns per mm of laser beam misalignment. Because the sign of the shift depends on the atom flow direction, the overall size of this effect is greatly reduced when using counter-propagating atomic beams as compared to a single beam, as long as the fluxes from the two ovens are balanced. With both atomic beams present, we measure a residual error no larger than 0.14~ns per mm of beam offset. By using the maximum drift ($100~\mu$m) in laser beam alignment observed during the 40-minute measurements, we determine an upper limit of $\pm0.014$~ns (0.05\%) for the uncertainty due to misalignment of the two laser beams.

To better understand this effect, we developed a simple model for a single atomic beam based on the following assumptions, which approximately match our experimental conditions:

\begin{enumerate}

\item
The excitation and ionization laser beams have Gaussian intensity profiles, propagate in the $z$ direction, and are centered on ($x = 0$, $y = 0$) and ($x_0$, $y_0$), respectively, as shown in Fig.~\ref{fig4-crossedbeams}.

\item
The atomic beam is well collimated, so that the atoms have negligible velocity components in the $y$- and $z$- directions. The velocity components in the $x$-direction are distributed according to the Maxwell-Boltzmann distribution for an atomic beam \cite{Ramsey1956},
\begin{equation}
    f(v)dv=\frac{v^3}{2\tilde{v}^4} \exp{ \left( \frac{-v^2}{2\tilde{v}^2} \right)}dv,
\end{equation}
where $\tilde{v}=\sqrt{kT/m}$ is a characteristic velocity.

\item
At the pump time $t=0$, an atom located at position ($x$, $y$) and moving with speed $v$ is excited with a probability proportional to the intensity of the excitation beam,
\begin{equation}
    I_1(x,y)=I_{10} \exp{\left( -\frac{2(x^{2}+y^{2})}{w_{1}^{~2}} \right)}
\end{equation} where $I_{10}$ is the on-axis intensity and $w_1$ is the $1/e^2$ beam radius.

\item
At the probe time $t > 0$, the same atom is located at ($x + vt$, $y$). If it is still in the excited state, as given by the probability $e^{-t/\tau}$, the atom is ionized with a probability proportional to the intensity of the ionization beam,
\begin{equation}
    I_2(x,y)=I_{20} \exp{\left( -\frac{2[(x+vt-x_0)^{2}+(y-y_0)^{2}]}{w_{2}^{~2}} \right)}
\end{equation}
where $I_{20}$ is the on-axis intensity and $w_2$ is the $1/e^2$ beam radius.

\end{enumerate}

The ion signal $S(t)$ is proportional to the product of the excitation intensity, the probability of an atom being in the excited state, and the ionization intensity, integrated over the probe volume and all atom velocities:
\begin{equation}
    S(t) \propto \iiint dx dy dz \int dv ~f(v) ~I_1(x,y) ~I_2(x,y) ~e^{-t/\tau}
\end{equation}
The volume integration is straightforward. If the characteristic distance $\tilde{v}t$ traveled by an atom during the measurement and the beam offset $x_0$ are both small compared to the laser beam dimensions, the integration over velocity can be carried out to give an ion signal with the approximate time dependence
\begin{eqnarray}\nonumber
   S(t) \propto e^{-t/\tau} e^{-2{x_0}^2/w^2} \left[ 1 + 3\sqrt{2\pi} \frac{x_0}{w} \left( \frac{\tilde{v}t}{w} \right) \right. \\
   \left. +~8 \left( \frac{4x_{0}^{~2}}{w^2} - 1 \right) \left( \frac{\tilde{v}t}{w} \right)^2 \right]
\end{eqnarray}
where $w=\sqrt{w_{1}^{~2}+w_{2}^{~2}}$ characterizes the sizes of the two laser beams.

The effect on the measured lifetime was found by using the known laser beam sizes in Eq.~8 to generate simulated ion signals. The simulated data were created in 1-ns time intervals over the range 0-500~ns and were fit using Eq.~1 to determine the lifetime. The results, determined separately for each atomic beam, are shown in Fig.~\ref{fig10-beamoffset} as solid lines. The slopes of these lines at $x = 0$ are $\pm 0.83(7)$~ns/mm, in reasonable agreement with the experimental data. When two counter-propagating beams are flowing simultaneously, and if the atom fluxes are balanced, the first order (in $\tilde{v}t/w$) effect vanishes for the total ion signal because the offset $x_0$ has opposite signs for the two beams. The atom fluxes are checked at the beginning of each run by separately observing the ion count rate from each cesium oven and are balanced to within about 20\%, which accounts for the residual lifetime shift of 0.14~ns/mm when both atomic beams are present. The lack of precise control of the atom fluxes is therefore the dominant source of uncertainty associated with this effect.

Even when the two oven fluxes are well balanced and the laser beams are perfectly aligned, Eq.~8 indicates that the ion counts are reduced by an amount $8(\tilde{v}t/w)^2$ from a purely exponential decay. The reduction is greater for longer delays, which shortens the apparent lifetime.  This second-order effect is additive and cannot be eliminated by using counter propagating-beams. To quantify the effect, we again used Eq.~8 to simulate the total ion signal for two counter-propagating atomic beams. We assumed the atom fluxes were perfectly balanced, so that the first-order term in Eq.~8 was canceled for the two beams. We also assumed the beams were perfectly aligned, so that the offset parameter $x_0$ was zero. As before, the simulated data were created in 1-ns intervals for the range 0-500~ns and were fit using Eq.~1 to determine the effect on the lifetime.  The results show a decrease in the lifetime of 0.18\%. We therefore applied a positive correction of 0.18\% to the extrapolated, zero-density lifetime shown in Fig.~\ref{fig9-radiationtrapping} to obtain a final value of 30.462~ns. The uncertainty in this correction is 0.02\% and arises in part from the uncertainties in the laser beam sizes and in the beam offset. The correction uncertainty also accounts for the slight temperature and, therefore, density dependence of the correction. Both the second-order correction and the first-order uncertainty due to flux imbalance could be greatly reduced by using higher laser intensities than were available for this experiment. Higher laser intensities would allow larger, more uniform beams to be used without sacrificing photo-ion signal, thereby reducing the corrective terms in Eq.~8.

\subsection{Hyperfine and Zeeman Quantum Beats}

\begin{figure}
\includegraphics[width=8.5cm]{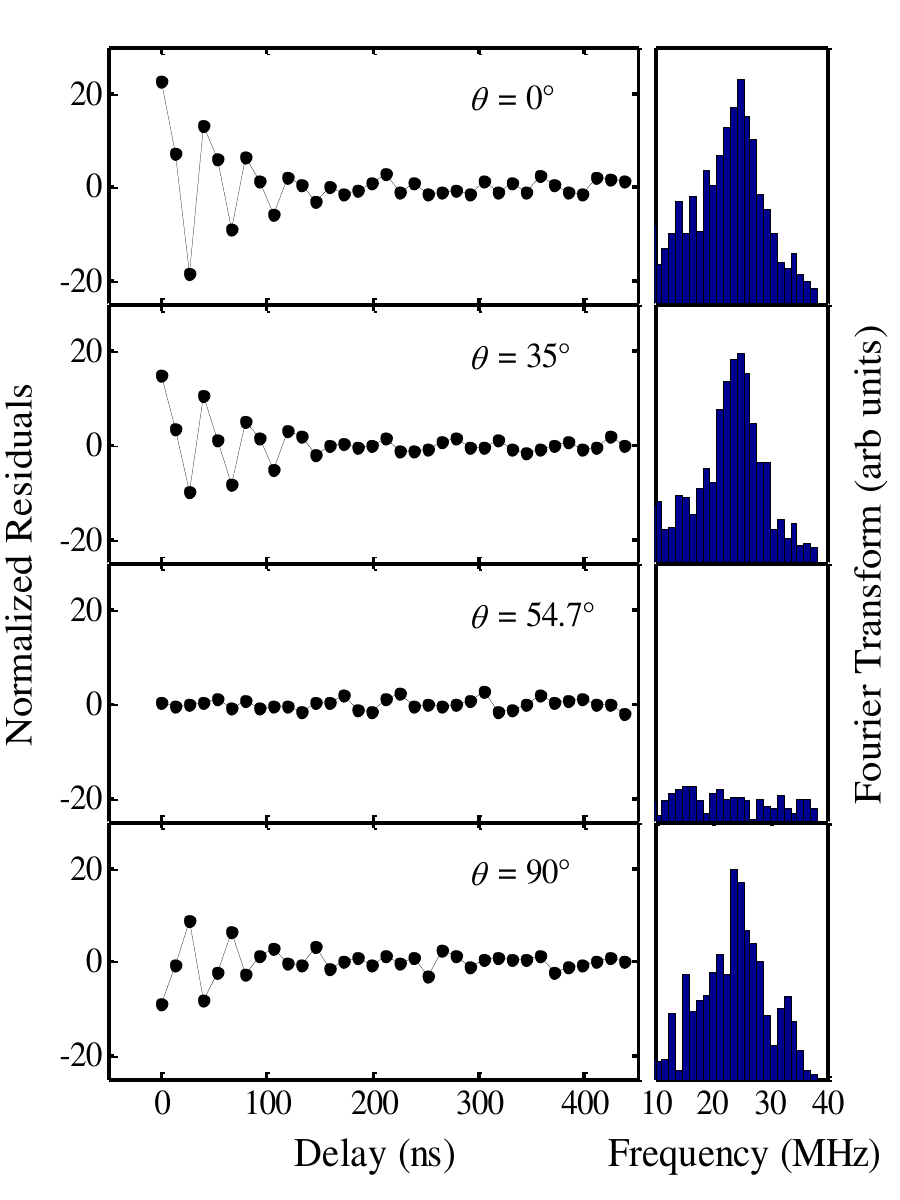}
\caption{\label{fig11-qbresiduals}   Fit residuals for data acquired at several polarization angles and illustrating the variation in the amplitude and phase of quantum beat oscillations, which are suppressed at the `magic angle' ($\theta_{m} = 54.7^\circ$). Each point represents the average residual of twenty individual lifetime measurements. For polarization angles deviating from $\theta_{m}$, the oscillations produce shifts in the measured lifetimes. The right-hand panel shows the Fourier transforms of the residuals.}
\end{figure}

Due to the broad bandwidth of our mode-locked laser, each excitation pulse coherently excites all four hyperfine states in the $6P_{3/2}$ manifold. The resulting quantum interference produces oscillations in the photo-ion signal known as quantum beats. Because these modulations are not incorporated in the fitting function (Eq.~1), the quantum beats are visible in the fit residuals (see Fig.~\ref{fig11-qbresiduals}) and produce a systematic error in the lifetime value extracted from the fit. The quantum beats vary in both amplitude and phase as the angle $\theta$  between excitation and ionization laser beam polarizations is changed. The oscillations are fully suppressed, to within the noise in the measurement, at the magic angle, $\theta \cong 54.7^{\circ}$. Consequently all data for our reported lifetime were acquired at the magic angle. Figure~\ref{fig11-qbresiduals} also shows the Fourier transforms of the residuals, which indicate an oscillation frequency of about 25~MHz. The actual quantum beat frequency is much higher, given the $6P_{3/2}$ hyperfine splittings of a few hundred MHz. Our sampling rate (76~MHz) is too slow to detect the quantum beats without aliasing.

\begin{figure}
\includegraphics[width=8.5cm]{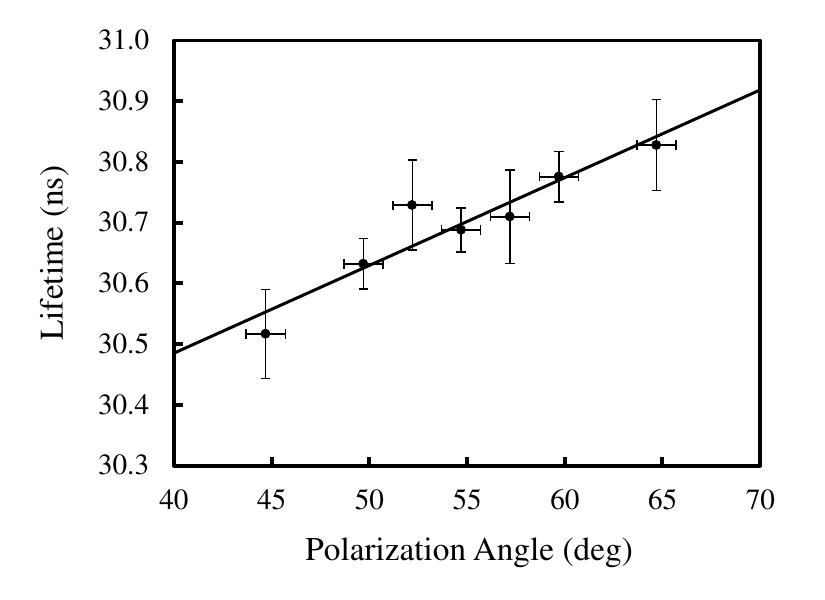}
\caption{\label{fig12-qbplot} The variation in the measured lifetime from changing the polarization angle $\theta$. Each point represents the average of 20 to 60 individual lifetime measurements. The slope of the linear fit is 0.014(5) ns/degree and is used to determine the experimental uncertainty due to hyperfine quantum beats.}
\end{figure}

To determine the experimental uncertainty arising from hyperfine quantum beats, we varied the polarization angle $\theta$ about the magic angle and measured the resulting lifetime shift, which was linear for the range of angles studied as shown in Fig.~\ref{fig12-qbplot}. Each point on this graph represents an average of 20-60 individual lifetime measurements, with the vertical error bars showing the statistical uncertainty. A weighted linear fit to the data yielded a slope of 0.014(5)~ns/degree. We take as the lifetime uncertainty this slope multiplied by the uncertainty in the polarization angle. The angular uncertainty arises from two sources: the measurement uncertainty for the polarization angle ($\pm 0.25^{\circ}$ as previously discussed) and an uncertainty introduced by possible birefringence in the vacuum chamber viewport. By measuring the polarization angle both before and after the viewport, we estimate a maximum uncertainty of $\pm 1.0^{\circ}$ for this effect.  The resulting lifetime uncertainty attributable to the total angular uncertainty is $\pm 0.018$~ns ($\pm 0.06\%$).

A related uncertainty is produced by the small ($\sim2\%$) amount of circularly polarized light present in the two beams. Assuming the worst case that 2\% of our data are acquired at $\theta = 0^{\circ}$, for which we see maximal quantum beats, there would be an additional lifetime uncertainty of $\pm 0.011$~ns ($0.04\%$). This value was estimated using 2\% of the lifetime shift predicted by the fit shown in Fig.~\ref{fig12-qbplot}, extrapolated to $\theta = 0^{\circ}$. The uncertainty is listed in Table~1 as `circular polarization.' The atoms in the cesium beams can also induce some rotation, as some of the light is off-resonant, but this effect is extremely small ($<4~\mu$rad) for our low cesium densities.

If a magnetic field is present, quantum beats can also arise from the coherent excitation of the Zeeman-split atomic levels. To minimize this effect, we used three pairs of mutually orthogonal Helmholtz coils (not shown in Fig.~\ref{fig2-setup}) to null the native magnetic field in the measurement region. As determined by a Hall probe, the native field has a magnitude of about 220~mG before nulling and is reduced to $\leq$ 20~mG when the nulling currents are applied to the coils. To estimate the associated measurement uncertainty, we separately measured the $6P_{3/2}$ lifetime with the native field present and with it nulled, and observed a 0.026-ns lifetime shift when the nulling currents are applied. Assuming the effect is linear in the magnetic field strength for small fields, there would be a small ($\leq$ 0.01\%) residual lifetime offset for a field strength of 20~mG, which we take as the uncertainty associated with Zeeman beats.

\subsection{Other Systematic Effects}

Because the mode-locked pulse interval $\Delta t$ directly determines the time scale for our measurement, the stability of the laser pulse train is a fundamental consideration. The pulse interval for our free-running laser can drift, for example, due to small changes in the cavity length with the ambient temperature, which is not controlled in this experiment. The laser pulse repetition rate was measured before and after each data run using an Agilent 53131A frequency counter having an RMS resolution of $<10^{-10}$ and an absolute time base uncertainty of $<4 \times 10^{-6}$. The input to the frequency counter was provided by a photodiode monitoring the excitation beam. The variation in the pulse repetition frequency during the measurements was consistently less than $6 \times 10^{-6}$. Combining this value with the time base uncertainty of the counter, we assign a measurement error of $<0.0003$~ns (0.001\%) for the laser pulse timing stability.

During selection of the excitation pulses, the unselected laser pulses are not fully eliminated due to the finite extinction ratio of the EOMs. These unwanted pulses provide a small additional excitation to the cesium atoms which is not accounted for by our simple exponential fitting function (Eq.~1). To minimize the resulting systematic error, each EOM was separately optimized by observing its output using a photodiode and oscilloscope, and adjusting its alignment and bias voltage to minimize the amplitude of the unselected pulses. Each EOM was found to have an extinction ratio better than 100:1 and, for the Conoptics model 350-160, the contrast ratio surpassed 300:1. The unselected pulses were not observable in the combined output of the EOMs and were below the noise level of the measurement, but we conservatively expect the effective contrast ratio for the three EOMs to exceed $10^5:1$ within about 50~ns of the selected pulse.

As pointed out by Young \emph{et al}.~\cite{Young1994}, the variation in the amplitude of the unselected pulses has a greater effect on the experimental uncertainty than the actual magnitude of these pulses. The variation in the pulse heights was the predominant source of error in that work and was termed `truncation error.' To evaluate the effect of these pulses on our measurement, we characterized the combined response of the three EOMs by the product of their individual transmissions. This combined response was used to simulate the excited state decay in the atomic beams. The simulated data were generated in 0.1-ns increments over the range 0-500~ns by summing, for each time step, the population decays created by forty preceding EOM leakage pulses, along with the principal decay from the selected excitation pulse. The resulting decay curve was fit using Eq.~1 and yielded a shift in the lifetime of less than 0.005\%, which we report as the uncertainty due to pulse selection.

Pulse pileup occurs when two or more ions reach the detector simultaneously, to within the dead time of the detector. Pulse pileup causes ions to be preferentially undercounted for higher counting rates, resulting in anomalously long lifetimes. Given the relatively long ion drift times ($\sim\mu$s), this effect will be small compared to measurements which detect fluorescence.   To minimize this source of error, the counting rates in our experiment are kept low, typically less than $1000$~s$^{-1}$.  Furthermore, the counting rates may be corrected using
\begin{equation}
    N' = \frac{N}{1 - N t_d}
\end{equation}
where $N$ and $N'$ are the measured and corrected counting rates, respectively, and $t_d$ is the detector dead time. For the dead time, we use the measured pulse width (20~ns) of the Channeltron detector, as the only way to `miss' an ion would be if two ions arrived within the same pulse width. Using Eq.~9 to correct each point on a decay curve produces a slight ($< 0.002\%$) negative shift in the fit lifetime, which we take as the pulse pileup uncertainty. This error is much smaller than observed in other experiments that use a time-to-analog converter, which typically can acquire only one count per cycle and one cycle may be $\sim1~\mu$s.

Finally, there is a potential error arising from the removal of ions from the atomic beam by the Channeltron. After the ions created by one pair of pump-probe laser pulses are removed, the cesium density is slightly reduced for the next pulse pair.  The ion depletion depends on the pump-probe delay and, in principle, could cause a systematic shift in the lifetime. The scale of this effect, however, is minuscule. Not only are the excitation and ionization fractions small (0.5\% and 0.2\%, respectively), but the ion-depleted zone of the atomic beam largely clears the measurement region during the 4-$\mu$s interval between measurements. We estimate the size of this effect to be less than 1 ppm, far smaller than any of the other systematic effects already considered.

\section{Discussion}

\begin{table}
\caption{\label{tab:table2} Comparison of precision lifetime measurements of the cesium $6P_{3/2}$ state.}
\begin{ruledtabular}
\begin{tabular}{lc}
Method  &  Lifetime (ns)  \\
\hline
Ultrafast excitation and ionization                     &  30.462(46)\footnote{This work; $^{b}$Ref.~\cite{Rafac1999}; $^{c}$Ref.~\cite{Young1994}; $^{d}$Ref.~\cite{Derevianko2002}; $^{e}$Ref.~\cite{Amini2003}; $^{f}$Refs.~\cite{Amiot2002, Bouloufa2007}}  \\
Fast-beam position-correlated photon counting           &  30.57(7)$^{b}$                              \\
Time-correlated single-photon counting                  &  30.41(10)$^{c}$                             \\
Van der Waals coefficient $C_{6}$                       &  30.39(6)$^{d}$                              \\
$6S_{1/2}$ static dipole polarizability                 &  30.32(5)$^{e}$                              \\
Photoassociation spectroscopy                           &  30.41(30)$^{f}$                             \\
\end{tabular}
\end{ruledtabular}
\end{table}

The uncertainties listed in Table~I were added in quadrature to determine an overall measurement uncertainty of 0.046~ns (0.15\%). Our final result for the Cs $6P_{3/2}$ lifetime is 30.462(46)~ns and is shown in Table~II, along with the results of other recent high-precision measurements. Our value falls between the two other direct measurements using position-correlated and time-correlated photon counting, differing by 0.35\% from that of Rafac \emph{et al}.~\cite{Rafac1999} and by 0.17\% from that of Young \emph{et al}.~\cite{Young1994}. The indirect measurement results include those obtained from the value of the van der Waals $C_{6}$ coefficient deduced from high-resolution Feshbach spectroscopy \cite{Derevianko2002}; from the Cs $6S_{1/2}$ static dipole polarizability measured in an atomic fountain experiment \cite{Amini2003}; and from high-resolution spectroscopy of photoassociated cold atoms \cite{Amiot2002, Bouloufa2007}. On average, the direct measurements give slightly higher lifetime values compared to the indirect techniques, although the measurement reported by Young \emph{et al}.~\cite{Young1994} aligns well with the results of the indirect techniques. The weighted average of all the results shown in Table~II is 30.421(26)~ns. The reader is referred to Rafac \emph{et al}.~\cite{Rafac1999} for a summary of additional, older measurements of the Cs $6P_{3/2}$ lifetime.

\begin{figure}[t]
\includegraphics[width=8.5cm]{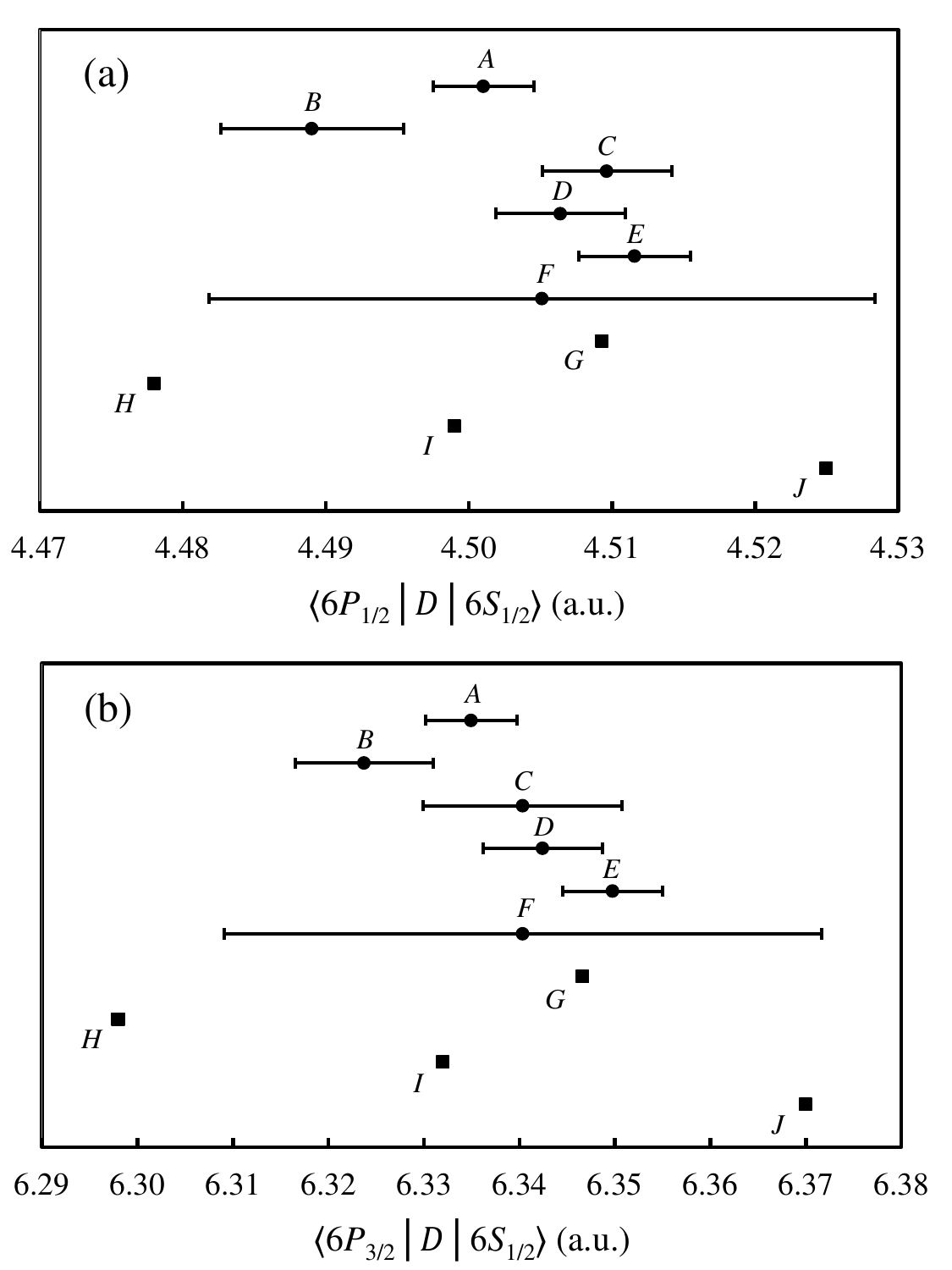}
\caption{\label{fig13-matrixelements} The reduced electric dipole matrix elements (a)~$\left\langle 6P_{1/2} \left\| D \right\| 6S_{1/2} \right\rangle$ and (b) $\left\langle 6P_{3/2} \left\| D \right\| 6S_{1/2} \right\rangle$ for cesium. The circles represent values extracted from the measurements shown in Table~II using Eq.~10, and the squares represent theoretical calculations. The labels correspond to the following references: (\emph{A})~This work; (\emph{B})~ref.~\cite{Rafac1999}; (\emph{C})~ref.~\cite{Young1994}; (\emph{D})~ref.~\cite{Derevianko2002}; (\emph{E})~ref.~\cite{Amini2003}; (\emph{F})~refs.~\cite{Amiot2002, Bouloufa2007}; (\emph{G})~ref.~\cite{Porsev2010}; (\emph{H})~ref.~\cite{Safronova1999}; (\emph{I})~ref.~\cite{Dzuba1989}; and (\emph{J})~refs.~\cite{Blundell1991, Blundell1992} }
\end{figure}

A measured lifetime can be used to calculate the reduced dipole matrix element, which is relevant for interpreting atomic parity non-conservation in Cs, using
\begin{equation}
    \frac{1}{\tau_{J}} = \frac{4}{3} \frac{\omega^{3}}{c^{2}} \alpha \frac{\left| \left\langle 6P_{J} \left\| D \right\| 6S_{1/2} \right\rangle \right| ^{2}}{2J + 1}
\end{equation}
where $\omega$ is the transition frequency, $c$ is the speed of light, $\alpha$ is the fine-structure constant, and $J$ is the angular momentum of the excited state. Our measured $6P_{3/2}$ lifetime gives a reduced dipole matrix element of $\left\langle 6P_{3/2} \left\| D \right\| 6S_{1/2} \right\rangle = 6.3349(48)$ in atomic units (a.u.). The $6P_{1/2}$ state, however, is more important for atomic parity nonconservation, as the corresponding matrix element directly enters into the amplitude of the PNC signal. It is possible to obtain the dipole matrix element for the $6P_{1/2}$ state by combining our $6P_{3/2}$ result with a high precision measurement of the relative line-strength ratio between these two states \cite{Rafac1998}. We obtain the result $\left\langle 6P_{1/2} \left\| D \right\| 6S_{1/2} \right\rangle = 4.5010(35)$ a.u. Figure~\ref{fig13-matrixelements} compares these matrix elements to those determined from other measurements \cite{Rafac1999, Young1994, Derevianko2002, Amini2003, Amiot2002, Bouloufa2007} and to a number of theoretical studies \cite{Safronova1999, Porsev2010, Dzuba1989, Blundell1991, Blundell1992}. The recent theoretical result for $\left\langle 6P_{1/2} \left\| D \right\| 6S_{1/2} \right\rangle$ by Porsev \emph{et al}.~\cite{Porsev2010} is a high precision \emph{ab initio} calculation utilizing the coupled cluster approximation. It is notable for the inclusion of valence triple excitations in the expansion of the cluster amplitude. This work yielded a weak charge for the cesium nucleus in good agreement with the standard model. The corresponding $\left\langle 6P_{3/2} \left\| D \right\| 6S_{1/2} \right\rangle$ result shown in Fig.~\ref{fig13-matrixelements}b for ref.~\cite{Porsev2010} was inferred from the $\left\langle 6P_{1/2} \left\| D \right\| 6S_{1/2} \right\rangle$ value using the relative line strength ratio between the two $6P$ states \cite{Rafac1998}. The other theoretical results \cite{Safronova1999, Dzuba1989, Blundell1991, Blundell1992} shown in Fig.~\ref{fig13-matrixelements} also employ the coupled cluster approximation and differ in the details of the included terms. Our results align most closely with the theoretical work of Dzuba \emph{et al.}~\cite{Dzuba1989}, with the results for both $\left\langle 6P_{1/2} \left\| D \right\| 6S_{1/2} \right\rangle$ and $\left\langle 6P_{3/2} \left\| D \right\| 6S_{1/2} \right\rangle$ agreeing with the calculations to within the measurement uncertainty. Our results differ by 0.18\% from the recent calculation of Porsev \emph{et al}.~\cite{Porsev2010}. The weighted averages of the experimental values shown in Fig.~\ref{fig13-matrixelements} are $\left\langle 6P_{1/2} \left\| D \right\| 6S_{1/2} \right\rangle$ = 4.505(2)~a.u. and $\left\langle 6P_{3/2} \left\| D \right\| 6S_{1/2} \right\rangle$ = 6.339(3)~a.u., falling approximately midway between the calculations of Dzuba \emph{et al.}~\cite{Dzuba1989} and Porsev \emph{et al.}~\cite{Porsev2010}.

Our technique achieves a measurement precision comparable to or surpassing other high-precision lifetime results, and could be further improved by a number of modest refinements. These include using active feedback to reduce the drift between the excitation and ionization laser beams to reduce the beam offset effect. Better polarization control would reduce the uncertainty from quantum beats, and a supplemental measurement of the atomic beam density would allow a better understanding of the radiation trapping effect. More significant improvements could be realized by using a narrow-linewidth CW laser to excite the atoms, while continuing to use the mode-locked laser for ionization. The excitation pulse train would be produced by shuttering the CW laser beam with EOMs, triggered by the synchronous signal from the regenerative amplifier. This modification would retain the precise timing of the mode-locked pulses, but would greatly reduce the largest error contributions listed in Table I. The CW laser would produce higher excited state fractions, enhancing the ion signal and improving the counting statistics. The higher excitation efficiency would also allow the excitation laser beam to be expanded and made more uniform, thereby reducing the beam offset effect discussed earlier (see Eq.~8). This modification would also reduce the uncertainty due to quantum beats, as we could selectively excite a single hyperfine state. Finally, it has the additional advantage of providing access to excited states which are far from the optimal wavelength of our regenerative amplifier ($\sim800$~nm), making the technique more generally applicable.

In conclusion, we have measured the cesium $6P_{3/2}$ excited state lifetime using ultrafast laser pulse excitation and ionization in counter-propagating thermal atomic beams. We have achieved an overall uncertainty of 0.15\%, which represents a 35\% improvement over the best previous direct measurement of this lifetime. This level of precision translates into an uncertainty of 0.08\% for the corresponding dipole matrix element, making the result useful for testing the validity of models used to interpret atomic parity violation experiments. In addition, we have discussed in detail the dominant sources of systematic error and offered a number of possible improvements to the experimental technique. We are currently implementing these upgrades in a second generation of the experiment and hope to achieve at least a factor of two improvement in precision. If successful, the anticipated precision may motivate improved theoretical calculations.

\begin{acknowledgments}
We gratefully acknowledge the Air Force Office of Scientific Research and the National Science Foundation (Grant Nos. 0758185, 1206128) for support of this work.
\end{acknowledgments}

\end{document}